\documentclass[aps,prd,twocolumn,showpacs,superscriptaddress,preprintnumbers]{revtex4-2}
 \usepackage[toc,page]{appendix}
 \usepackage[utf8]{inputenc}
 \usepackage{amsmath,amsthm,amssymb,mathtools,physics,cancel,mhchem,tikz}
 \usepackage{graphicx}
 \usepackage{dcolumn}
 \usepackage[hidelinks]{hyperref}
 \hypersetup{colorlinks=true,
     linkcolor=blue,
     filecolor=blue,
     urlcolor=blue,
     citecolor=blue,
     }

 \usepackage{ulem}
 \usepackage{adjustbox}
 \usepackage{gensymb}
 \usepackage{breqn}
 \usepackage{braket}
 \usepackage{enumitem}
 \usepackage{gensymb}
 \graphicspath{{./figures/}}

\begin{document}
\begin{flushright}
MI-HET-827
\end{flushright}
\title{Prospects for Light Dark Matter Searches at Large-Volume Neutrino Detectors
}

\begin{abstract}
We propose a new approach to search for light dark matter (DM), with keV-GeV mass, via inelastic nucleus scattering at large-volume neutrino detectors such as Borexino, DUNE, Super-K, Hyper-K, and JUNO.
The approach uses inelastic nuclear scattering of cosmic-ray boosted DM, enabling a low-background search for DM in these experiments.
Large neutrino detectors, with higher thresholds than dark matter detectors, can be used since the nuclear deexcitation lines are $O(10)$ MeV.
Using a hadrophilic dark-gauge-boson-portal model as a benchmark, we show that the nuclear inelastic channels generally provide better sensitivity than the elastic scattering for a large region of light DM parameter space.
\end{abstract}

\author{Bhaskar Dutta}
\email{dutta@physics.tamu.edu}
\affiliation{Mitchell Institute for Fundamental Physics and Astronomy$,$ Department of Physics and Astronomy$,$ \\Texas A\&M University$,$ College Station$,$ Texas 77843$,$ USA}

\author{Wei-Chih Huang}
\email{s104021230@tamu.edu}
\affiliation{Mitchell Institute for Fundamental Physics and Astronomy$,$ Department of Physics and Astronomy$,$ \\Texas A\&M University$,$ College Station$,$ Texas 77843$,$ USA}

\author{Doojin Kim}
\email{doojin.kim@usd.edu}
\affiliation{Mitchell Institute for Fundamental Physics and Astronomy$,$ Department of Physics and Astronomy$,$ \\Texas A\&M University$,$ College Station$,$ Texas 77843$,$ USA}
\affiliation{Department of Physics$,$ \\University of South Dakota$,$ Vermillion$,$ South Dakota 57069$,$ USA}

\author{Jayden~L.~Newstead}
\email{jnewstead@unimelb.edu.au}
\affiliation{ARC Centre of Excellence for Dark Matter Particle Physics$,$ School of Physics$,$ \\The University of Melbourne$,$ Victoria 3010$,$ Australia}

\author{Jong-Chul Park}
\email{jcpark@cnu.ac.kr}
\affiliation{Department of Physics and Institute of Quantum Systems (IQS)$,$ \\Chungnam National University$,$ Daejeon 34134$,$ Republic of Korea}

\author{Iman Shaukat Ali}
\email{ishaukatali@student.unimelb.edu.au}
\affiliation{ARC Centre of Excellence for Dark Matter Particle Physics$,$ School of Physics$,$ \\The University of Melbourne$,$ Victoria 3010$,$ Australia}

\maketitle

\noindent {\bf Introduction.} The nature of dark matter (DM) is still a mystery although
decades have passed since the first evidence of its existence was found.
To date, DM has only been detected via gravitational interactions.
Identifying the particle nature of DM through non-gravitational couplings is the current subject of diverse international experimental efforts, e.g., direct detection of DM utilizing nuclear/electronic recoils, indirect detection from DM annihilation/decay, and direct DM production at colliders.

The paradigm of weakly interacting massive particles (WIMPs) for DM is favored due to its insensitivity to underlying dark-sector scenarios.
Being well-motivated and popular DM candidates, WIMPs have become the primary focus of various DM search efforts (see, e.g., Refs.~\cite{Arcadi:2017kky,Elor:2015bho,Schumann:2019eaa,Cooley:2022ufh}).
The lack of a conclusive WIMP signal has increasingly motivated alternative possibilities.
Among them, models of (MeV-scale) light DM involving new feebly-interacting (MeV-scale) light mediators have received growing attention.
Light mediators are required as many light DM models involve large scattering cross-sections.
The limit on scattering cross-sections imposed by contact interactions is the geometric size of the nucleus.
However, light mediators allow for longer-range interactions that exceed this limit~\cite{Digman:2019wdm}.
These models can explain the DM content of the universe without invoking WIMPs, while keeping the thermal nature of DM.
However, the experimental detection of such DM candidates is challenging as the aforementioned search efforts are designed and optimized for heavier (GeV-TeV scale) WIMP-type DM.

In contrast, fixed-target experiments utilizing high-intensity particle beams are better suited to probing light DM, as the beam energy and intensity are large enough to produce feebly-interacting MeV-range DM particles through the decay of a mediating particle~\cite{Dutta:2019nbn,Dutta:2020vop,Batell:2022xau}.
Moreover, the relativistic nature of the produced DM enables the detection of DM signals without ultra-low threshold detectors.
These experiments mainly use the elastic nuclear scattering channel, especially in the search for hadrophilic DM.
However, it has been shown that the {\it in}elastic nuclear scattering channel can achieve significantly better reach compared to the elastic channel in fixed-target experiments, for example, beam-dump neutrino experiments~\cite{Dutta:2023fij}.
This is because the line-like photon signals from the associated nuclear deexcitation enjoy both more energy and a higher signal-to-background ratio than the elastic channel.

Inspired by the observation made in the beam-produced DM search, in this Letter, we take advantage of and apply nuclear inelastic scattering for the search for cosmic-ray boosted DM signals.
Nuclear excitation energies of detection targets in large-volume detectors, e.g., argon, oxygen, and carbon, are in the MeV range.
Therefore, it is kinematically impossible for the non-relativistic population of ambient DM to leave detectable signatures through the inelastic channel.
On the other hand, various mechanisms of ``boosting'' ambient DM itself or its subcomponents, in the present universe, have been proposed~\cite{DEramo:2010keq,Belanger:2011ww,Agashe:2014yua,Kong:2014mia,Kim:2016zjx,Bringmann:2018cvk,Ema:2018bih,Jho:2021rmn} such that a relativistic DM signal could be detected in the inelastic channel.

The DM that we consider here is boosted by cosmic rays~\cite{Bringmann:2018cvk,Ema:2018bih,Jho:2020sku,Jho:2021rmn,PhysRevD.104.076020,PhysRevD.101.116007,Bell:2023sdq,Guha:2024mjr}, scattering {\it in}elastically in a neutrino detector, and creating a line signal of $O(10)$~MeV.
We choose this boosting method for illustration, noting that the inelastic channel can be used for other types of boosted DM.
More specifically, we focus on light-dark-sector models where the DM and dark-sector mediator masses are in the keV-to-GeV range.
The cosmic-ray boosting employs the same interactions that allow for detection through DM-nucleus scattering.
Since low-threshold detectors are not essential for these $O(10)$~MeV line signals, we can make use of present and future large-volume neutrino detectors such as Borexino, DUNE, Super-K, Hyper-K, and JUNO.
We demonstrate that for large neutrino detectors the inelastic nuclear scattering channels (characterized by line signals) can achieve much better sensitivity compared to the elastic channel.
Moreover, this method probes a much larger region of parameter space as compared to searches at fixed-target experiments, which require that the DM be lighter than the mediating particle.

\medskip

\noindent {\bf DM-nucleus interaction.}
We consider a dark-sector model where a dark gauge boson, $A'$, couples to both the DM, $\chi$, and the nucleus.
As a concrete example, we consider the case where $A'$ couples to the SM quarks, resulting in the following operators:
\begin{equation}
        \mathcal{L} \supset g_D A'_\mu \bar \chi \gamma^\mu \chi + \epsilon Q_b A'_\mu \bar q \gamma^\mu q\,,
    \label{eq:int}
\end{equation}
where $\epsilon$ is nucleon coupling, the dark coupling constant is taken to be $g_D=\sqrt{2\pi}$, and $Q_b$ parameterizes the baryon number of quark species $q$.
\footnote{The kinetic mixing ($\varepsilon$) arises at the loop level and is given by $\varepsilon\sim {e N_f \epsilon \over 16 \pi^2}$ where $N_f$ is the number of quark flavors.}
Scenarios of this sort can arise together with various new gauge bosons, e.g., $U(1)_{T3R}$~\cite{Dutta:2019fxn}, $U(1)_{B-3L_{\mu,\tau}}$~\cite{Farzan:2016wym}, and $U(1)_B$~\cite{deNiverville:2015mwa}.
In this work, we will only consider DM-quark interactions to explore inelastic and elastic \textit{nuclear} scattering.
DM-electron interactions, if present in particular models, could also be used as a complementary probe~\cite{Cappiello:2018hsu,Ema:2018bih,Jho:2021rmn}.

Depending on the particular model of interest, the dark-gauge-boson mass required to produce a viable signal via cosmic-ray boosting may be constrained by direct searches~\cite{Bell:2023sdq, Guha:2024mjr}.
For example, in the case of coupling to an anomalous current in a $U(1)_B$ model, couplings to the $A'$ are constrained by meson decays~\cite{Dror:2017ehi}.
On the other hand, anomaly-free examples have more available parameter space.
We also note that CRDM with a light mediator can be constrained by cosmological considerations, e.g., Big Bang Nucleosynthesis for mass $\lesssim 5$ MeV~\cite{Krnjaic:2019dzc}.

The experimental signature under consideration is the deexcitation photon from an excited nucleus produced by nuclear inelastic scattering.
Irreducible backgrounds are produced by inelastic neutrino-nucleus scattering, for which we take solar, atmospheric, diffuse supernova, and geo-neutrinos into account as sources.

\noindent {\it i) \underline{Boosting DM with cosmic rays}.}
Due to the couplings in Eq.~\eqref{eq:int}, (light) DM can be boosted from the scattering of energetic cosmic rays throughout the galaxy before they come to the detectors; in particular, we consider protons and helium.
We follow the framework of Ref.~\cite{Bringmann:2018cvk} in calculating the flux of DM boosted by cosmic-ray particles.
While it may be interesting to investigate the production of CRDM via inelastic processes~\cite{Alvey:2019zaa}, here we consider elastic upscattering only.

The differential elastic scattering cross-section of DM boosted by cosmic-ray species $i$ is
\begin{align}
    \label{eq:CRDM_xsec}
    &\,\,\,\,\frac{d\sigma^{\mathrm{el}}_{\chi i}}{dT_\chi} = g_D^2 \epsilon^2 A^2 F^2 \\ \nonumber
    &\times \frac{ m_\chi(m_i+T_i)^2-T_\chi \left\{ (m_i+m_\chi)^2+2 m_\chi T_i\right\} + m_\chi T^2_\chi }{4\pi(2m_\chi T_\chi + m_{A^\prime}^2)^2(T_i^2 + 2m_i T_i)}\,,
\end{align}
where $m_{\chi/i}$, $A$, $F$, and $T_{\chi/i}$ are the mass of DM/incident cosmic-ray $i$, the mass number of the cosmic ray, the form factor (e.g., dipole form factor for protons and helium, and Helm form factor for heavier nuclei), and the kinetic energy of DM/cosmic-ray, respectively.
The initial DM is assumed to be at rest in the galactic frame.
The collision rate $\Gamma_{i\to \chi}$ of cosmic-ray species $i$ with DM in an infinitesimal volume $dV$ is~\cite{Bringmann:2018cvk}
\begin{equation}
    \label{eq:col_rate}
    \frac{d^2\Gamma_{i\to \chi}}{dT_idT_\chi} = \frac{\rho_\chi}{m_\chi} \frac{d\sigma^{\mathrm{el}}_{\chi i}}{dT_\chi} \frac{d\Phi^{\rm{LIS}}_i}{dT_i} dV\,,
\end{equation}
where $d\Phi^{\rm{LIS}}_i / dT_i$ is the local interstellar flux of cosmic-ray species $i$ and $\rho_\chi=0.3~{ \rm GeV}\cdot{\rm cm}^{-3}$ is the DM density near the Earth.
The integration of Eq.~\eqref{eq:col_rate} over $dV$ and $dT_i$ essentially yields the cosmic-ray boosted DM (CRDM) flux
\begin{align}
    \label{eq:CRDM}
    \frac{d\Phi_\chi}{dT_\chi} &=\int_V\!dV\int_{T_i^{\rm{min}}}\!\!\!\!dT_i \,\, \frac{d^2\Gamma_{i\rightarrow\chi}}{dT_idT_\chi}\nonumber \\
    &= D_{\rm eff}\,\frac{\rho_\chi}{m_\chi}\, \sum_i \int_{T_i^{\rm{min}}}\!\!\!\!dT_i\,\frac{d\sigma^{\mathrm{el}}_{\chi i}}{dT_\chi}\,\frac{d\Phi^{\rm{LIS}}_i}{dT_i}\,,
\end{align}
where, for a 1-kpc radius sphere centered on the Earth, $D_{\rm eff} = 0.997$~kpc is the effective diffusion zone (which takes into account the varying DM density)~\cite{Bringmann:2018cvk}.
See Supplement for the CRDM flux for various $m_\chi$ and $m_{A'}$.

\noindent {\it ii) \underline{Inelastic nuclear scattering of CRDM}.}
Borrowing from the treatment of relativistic DM-nucleus inelastic scattering for fixed-target experiments, it was shown that in the long-wavelength limit, Gamow-Teller (GT) transitions are the dominant contribution to the total inelastic cross-section ($\gtrsim$ 90\%)~\cite{dutta:2022tav}.
Thus, we can approximate the cross-section of $\chi+N\rightarrow\chi+N^*$, by considering only GT transitions~\cite{dutta:2022tav}:
\begin{align}
    \label{eq:DM_xsec}
    \frac{d\sigma^{\mathrm{inel}}_{\chi N}}{d\cos\theta} &= \frac{2\epsilon^2 g_D^2 {E'}_\chi {p'}_\chi}{(2m_T E_R + m_{A'}^2 -\Delta E^2)^2} \frac{1}{2\pi} \frac{4\pi}{2J+1} \\
    &\times \sum\limits_{s_i,s_f} \vec{l} \cdot \vec{l}^* \frac{g_A^2}{12\pi}
    |\langle J_f|| \sum_{i=1}^A \frac{1}{2}\hat{\sigma_i} \hat{\tau_0}|| J_i\rangle|^2\,, \nonumber
\end{align}
where $E_\chi' / p_\chi'$, $m_T/J$, $E_R$, $\Delta E$, and $g_A=1.27$ are the outgoing $\chi$ energy/momentum, target nucleus mass/spin, nuclear recoil energy, excitation energy, and axial form factor, respectively.
$|\langle J_f|| \sum_{i=1}^A \frac{1}{2}\hat{\sigma_i} \hat{\tau_0}|| J_i\rangle|^2$ in the second row describes the GT transition strength.
$l_\mu=\bar \chi \gamma^\mu \chi$ is the DM vector current and summing them over spins is
\begin{equation}
    \sum\limits_{s_i,s_f} \vec{l} \cdot \vec{l}^* = 3-{1\over E_\chi E_\chi^\prime} \left[ {1\over2} \left({p_\chi}^2 + {p_\chi^\prime}^2-2m_T E_R \right) +{3m_\chi^2\over4} \right]\,.
\end{equation}

\noindent {\it iii) \underline{GT transition strengths}.}
The calculation of the cross-section in Eq.~\eqref{eq:DM_xsec} requires knowledge of the GT strength for the relevant nuclear transitions of the target.
This can be derived from experimental results or computed via, e.g., the nuclear shell model.

For the argon target, we compute the GT strengths and excitation energies, using the large-scale shell model code BIGSTICK~\cite{Johnson:2018hrx,Johnson:2013bna}, employing the {\it SDPF-NR} interaction \cite{Prados:PRC2007,Nowacki:PRC2009,Nummela:PRC2001}, and considering a single isotope, $^{40}$Ar only.
We do truncation to $^{40}$Ar to reduce computational workload.
We refer to Supplement for details.
The experimental data in Ref.~\cite{Tornow:2022kmo} shows that there is a scale factor of 0.16 for the strength of $^{40}$Ar in $\sim 4-12$ MeV energy range, but does not provide the strengths of individual lines.
Thus, the shell model calculation is inevitable and its accuracy can be improved by this scale factor.

The GT strengths for carbon and oxygen targets are derived from experimental measurements.
For carbon, we employ the measurements of neutrino-nucleus scattering made by KARMEN~\cite{Maschuw:1998qh,KARMEN:1991vkr,Suliga:2023}, which found the GT strength of the ground-state-to-$^{12}\mathrm{C}^*(15.1$ MeV) transition to be 0.255.
For oxygen, we take the $^{16}$O magnetic dipole measurements~\cite{KUCHLER1983473,TILLEY19931}, from which we derive the GT strengths (in the long-wavelength limit) through a scaling relationship~\cite{PhysRevLett.93.202501}.
Refs.~\cite{TILLEY19931,PhysRevLett.43.117,Rangacharyulu:1983obd} report the experimental measurements of $^{16}$O branching ratio ${\Gamma_{N^*\to N\gamma} \over \Gamma_{\rm total}}$ in the order of $\mathcal{O}(10^{-5}-10^{-4})$,
\footnote{They are $1.56\times 10^{-5}$, $1.68\times 10^{-4}$, $1.79\times 10^{-4}$, and $4.42\times 10^{-5}$ for 13.664 MeV, 16.22 MeV, 17.14 MeV, and 18.79 MeV states, respectively.} (whereas those of $^{12}$C and $^{40}$Ar are $\sim1$).
The low branching ratios allow for relatively less signal and (neutrino-induced) background with oxygen, which can be compensated by large detector volume.

\medskip

\noindent {\bf Benchmark detectors.}
Conventional DM direct detection experiments aim to measure the nuclear recoil induced by the elastic scattering of WIMPs off the detector target.
With ``typical'' WIMP masses of $\mathcal{O}(10-1,000)$~GeV, detectors require keV-range energy thresholds due to the non-relativistic nature of halo DM.
Ultra-low threshold methods, such as cryogenic bolometers~\cite{SuperCDMS:2020aus}, superfluid helium~\cite{Schutz:2016tid,Knapen:2016cue,Maris:2017xvi}, or the Migdal effect~\cite{Dolan:2017xbu, SuperCDMS:2023sql}, are required to search for WIMPs below GeV.
The relativistic CRDM component overcomes this limitation and can deposit $>$MeV of energy even for WIMP masses below MeV.
However, the CRDM flux is steeply falling with energy, only allowing CRDM to be detectable for relatively large cross-sections.
This also gives rise to a situation where large-volume neutrino detectors, with much higher energy thresholds than direct detection experiments, can be the most sensitive probe of the CRDM flux (depending on the type of scattering and model parameters)~\cite{Bringmann:2018cvk,Ema:2020ulo,Super-Kamiokande:2022ncz,Bell:2023sdq}.
Given these considerations, kton-scale neutrino detectors with MeV-range energy thresholds are also well suited to search for CRDM via nuclear inelastic scattering.

In this context, we consider inelastic scattering in the current detectors, Borexino and Super-K, as well as the future detectors, JUNO, Hyper-K and DUNE.
Borexino~\cite{Borexino:2008gab, Borexino:2000uvj} ran from 2007 to 2021 with a 100-ton organic liquid scintillator, achieving a 5.6-year equivalent data exposure.
Although it was designed for solar neutrino detection, its data could be reinterpreted in terms of a DM search due to its large size and low energy threshold.
Here we provide an analysis that would demonstrate the sensitivity attainable if such an analysis were performed.
Super-K~\cite{Super-Kamiokande:2002weg,Super-Kamiokande:2022ncz,Suzuki:2019jby} has a 22.5-kton fiducial mass of water, operating for 20 years.
A proper recast of Super-K limits is nontrivial as they have not carried out similar searches for low-energy gamma rays ($\lesssim30$ MeV).
JUNO~\cite{JUNO:2022103927, JUNO:2015zny, JUNO:2015sjr} is a 20-kton organic liquid scintillation detector, expected to begin operations in 2024.
Hyper-K~\cite{Hyper-Kamiokande:2018ofw,Hyper-Kamiokande:2022smq} will be a 188-kton water Cherenkov detector, and is planning to begin operation in 2027.
DUNE~\cite{DUNE:2015,DUNE:2020lwj,DUNE:2020ypp} will be a 40-kton liquid-argon time projection chamber (TPC) detector and will start cosmic-origin data collection later this decade.
Additionally, for comparison purposes, we also consider elastic scattering in the above detectors (with appropriate thresholds) as well as in LZ~\cite{LZ:2022lsv} and DarkSide-20k~\cite{DarkSide-20k:2017zyg}.
We summarize the specifications of these detectors in Table~\ref{tab:spec}.

\begin{table}[t]
    \centering
    \caption{Key specifications of the detectors.
    The top (bottom) section is used in the inelastic (elastic) analyses.
    Energy resolution $E_{\rm res}$ and range of deposited energy $E_{\rm range}$ are in MeV.
    The time for Borexino means the amount of actual (fiducial) exposure time.}
    \hspace*{-0.85cm}
    \hspace{4mm}
    \resizebox{\columnwidth}{!}{
    \begin{tabular}{c|c c c c c c}
    \hline \hline
    Experi- & Target & Mass & Time & $E_{\rm res}$ & $E_{\rm range}$ & Back-\\
    ment &  & (kton) & (year) & (MeV) & (MeV) & ground\\
    \hline
    Super-K & water & 22.5  & 20  & 2  & 10 - 22 & $2.9\!\times\!10^{-3}$ \\
    Hyper-K& water & 188  & 10  & 2  & 10 - 22 & $1.2\!\times\!10^{-2}$ \\
    DUNE & Ar & 40  & 10  & 2  &  7 - 15 & 167.5 \\
    JUNO & ${\rm CH}_2$ & 20  & 10  & -  & 15.1 & 47\\
    Borexino & ${\rm CH}_2$ & 0.1 & 5.6 & - & 15.1 & 0.13\\
    \hline
    DUNE & p,n & 40  & 10  & -  & 50 - 1270 & 550 \\
    JUNO & p & 20  & 10  & -  & 15 - 100 & 1420 \\
    Borexino & p & 0.1 & 1.22  & -  & 12.3 - 24.6 & 84\\
    LZ & Xe & 0.0055 & 0.167  & -  &  (1$\,$-$\,$68)$\cdot10^{\mbox{-}3}$ & 1.4 \\
    DarkSide & Ar & 0.02 & $5$  & - & 0.03 - 0.1 & 1.6 \\
    \hline \hline
    \end{tabular}
    }
    \label{tab:spec}
\end{table}

\medskip

\noindent {\bf Background considerations.}
The main backgrounds to the above-described CRDM signals arise from the elastic/inelastic neutral-current scattering of neutrinos. As mentioned earlier, we consider solar, atmospheric, diffuse supernova, and geo-neutrino sources \cite{Suzuki:solar,Suzuki:atm,SRAMEK2013356,PhysRevD.99.093009,Vitagliano:2019yzm}.
The solar and atmospheric fluxes make the dominant contribution to the background.

\begin{figure}[t]
    \centering
    \includegraphics[width=\columnwidth]{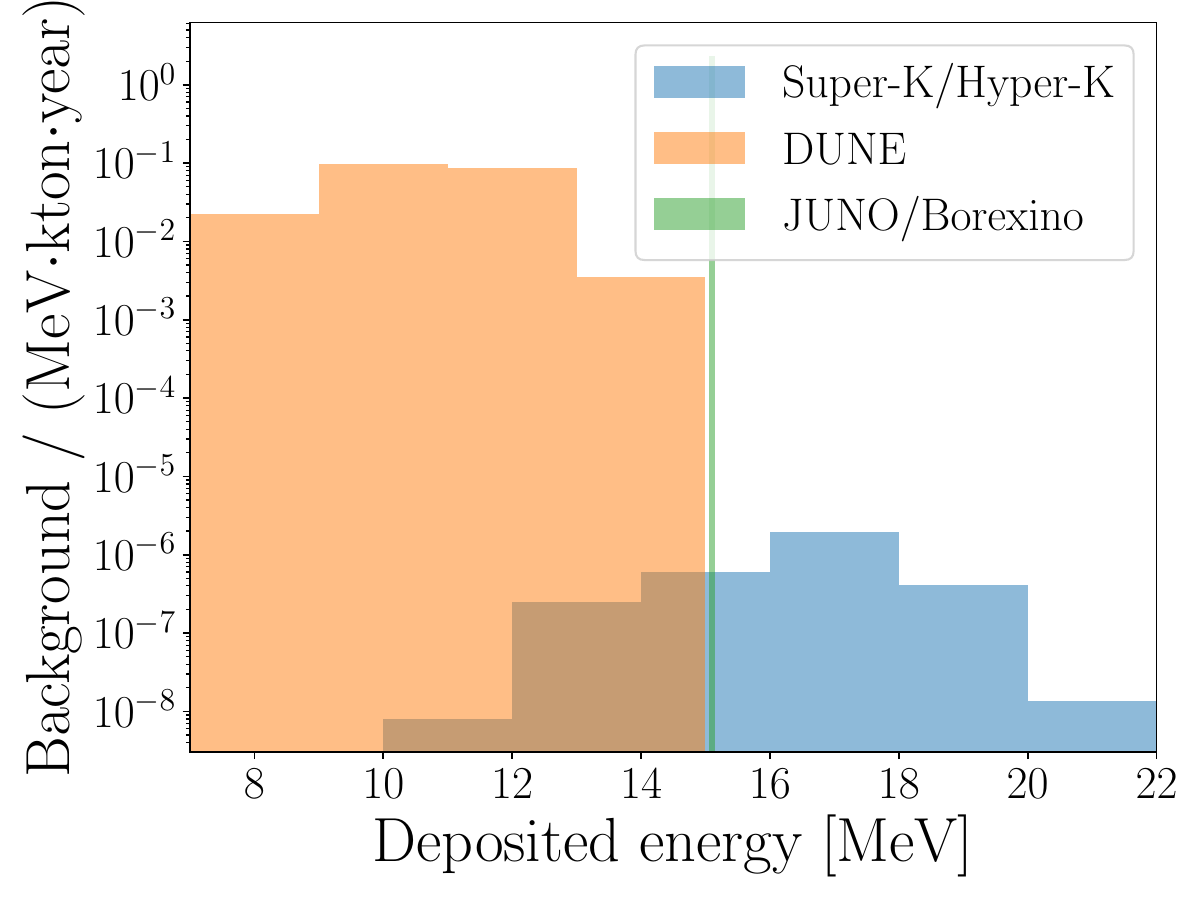}\\
    \caption{Estimated background rates to inelastic signals for the detectors considered.
    Both Super-K and Hyper-K use water Cherenkov detectors, so their estimated backgrounds per (kton-year) are identical. Similar for JUNO and Borexino.
    }
    \label{fig:bkg}
\end{figure}

Using the neutrino fluxes and the inelastic scattering cross-section, one can estimate the neutrino-induced backgrounds for our benchmark detectors: Super-K, Hyper-K, and DUNE.
To maintain the coherency of a nucleus in the scattering, an upper bound of nuclear recoil momentum, $q<100$ MeV, is applied in our analysis.
The background (similar to the signal) is proportional to nuclear GT strengths.
We show our background estimates in Fig.~\ref{fig:bkg} with the detector energy resolutions and ranges listed in Table~\ref{tab:spec}.

For JUNO and Borexino, Ref.~\cite{Suliga:2023} considered inelastic $\nu$-$^{12}$C scattering of atmospheric neutrinos in JUNO, finding a total of 20 events, which consist of 16 neutrino-induced nuclear inelastic scattering events and 4 background events such as
nuclear inelastic events by solar and DSNB neutrino, and continuum from elastic $\nu$-$e$ scattering (ignorable), for an 85 kton-year exposure.
Therefore, we expect 47 (0.13) background events at JUNO (Borexino) for its projected (completed) 200 (0.56) kton-year exposure.

\medskip

\noindent {\bf Analyses and results.}
We report our results in both nucleus elastic and inelastic (GT) scattering channels of CRDM.
The earlier efforts in the search for CRDM are done in the elastic scattering channels in the experiments such as Super-K ~\cite{Super-Kamiokande:2022ncz}, PandaX-II~\cite{PandaX-II:2021kai}, and CDEX-10~\cite{CDEX:2022fig}.
In the elastic scattering analysis, we consider data from LZ~\cite{LZ:2022lsv} and Borexino~\cite{Borexino:2013bot}, and set projections for future detectors, JUNO~\cite{Chauhan:2021fzu}, DarkSide-20k~\cite{DarkSide-20k:2017zyg}, and DUNE~\cite{Berger:2019ttc}.
Our bounds are placed by calculating the coupling $\epsilon$ that produces the 90\% confidence limit (CL) on additional events from CRDM, in the range of accessible nuclear recoil energies.

The LZ experiment is the current largest liquid-xenon TPC.
Using preliminary data of 60 days and a fiducial mass of 5.5~ton, we consider the region below the median nuclear recoil band and apply an additional 50\% to the detection efficiency to account for this nuclear recoil cut \cite{LZ:2022lsv}.
One event was observed in this region while 1.4 events were expected, which corresponds to an upper limit of 2.3 CRDM events.\footnote{Both XENON1T~\cite{XENON:2018voc} and XENONnT~\cite{XENON:2023cxc} have collected nearly the same amount of data as LZ, resulting in similar sensitivity reaches.}
For Borexino, we follow the analysis in Ref.~\cite{Borexino:2013bot}, and consider events in the electron-equivalent range of $E_e =4.8 - 12.8$ MeV (corresponding to proton recoil energies in the range $E_R=12.3 - 24.6$ MeV).
The total observed background in this region was 84 events, while 71 were expected.
We only consider hydrogen as a target, ignoring scattering on carbon as it is kinematically suppressed.
Building on its predecessor DarkSide-50, DarkSide-20k will be a liquid-argon TPC with a fiducial mass of 20 tonnes.
Assuming an exposure time of 5 years, we consider argon recoils in the range $E_R =30 - 100$ keV, where a total of 1.6 background events are expected \cite{DarkSide-20k:2017zyg}.
The JUNO projection assumes elastic scattering off hydrogen in the proton recoil range  $E_R = 15 - 100$ MeV \cite{Chauhan:2021fzu}. 142/(20~kton-year) background events are expected in this region. We set a total exposure time of 10 years.
Finally, DUNE elastic scattering projections are calculated by searching for proton and neutron knockout events in the energy range 50 MeV $-$ 1.27 GeV.
We estimate 55 background scattering events per 40~kton-years from atmospheric neutrino neutral-current interactions \cite{Berger:2019ttc}.

\begin{figure*}[t]
    \centering
    \includegraphics[width=2\columnwidth]{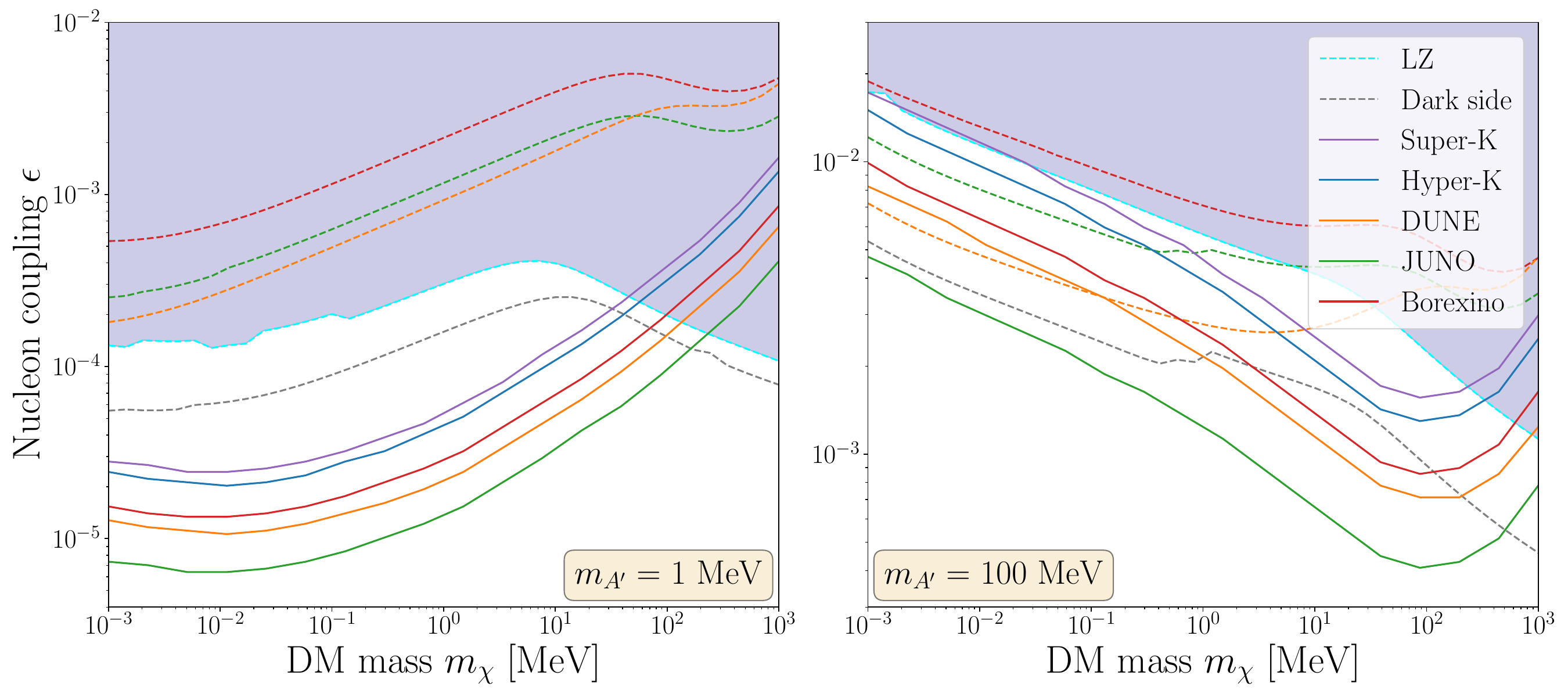}
    \caption{90\% CL sensitivity reaches in the search for ambient DM through the CRDM flux in the $(m_\chi, \epsilon)$ plane.
    The left (right) panel shows the results with $m_{A'}=1$~MeV ($m_{A'}=100$~MeV).
    Solid lines denote the nuclear inelastic channels, while dashed lines represent nuclear elastic channels.
    The shaded region shows the current bounds while all other curves are projected/attainable sensitivities.
    }
    \label{fig:sen}
\end{figure*}

In the nuclear inelastic channel, the CRDM flux $\frac{d\Phi_\chi}{dE_\chi}$ in Eq.~\eqref{eq:CRDM} and the inelastic scattering cross-section in Eq.~\eqref{eq:DM_xsec} yield the expected number of signal events, $N_\chi$ as follows:
\begin{equation}
    N_\chi = N_T \Delta t \int \sigma^{\mathrm{inel}}_{\chi N}(E_\chi) \frac{d\Phi_\chi}{dE_\chi} dE_\chi\cdot\frac{\Gamma_{N^*\to N\gamma}}{\Gamma_{\rm total}}\,,
\end{equation}
where $N_T$ and $\Delta t$ and are the number of (scattering-target) nuclei in the detector and the detector operation time, respectively.

Based on the above formalism and the masses and exposure times of Table~\ref{tab:spec}, we first plot expected 90\% CL sensitivity reaches in the $(m_\chi, \epsilon)$ plane in Fig.~\ref{fig:sen}.
The solid (dashed) lines are projected reaches in the inelastic (elastic) channels, while the shaded regions are current bounds.
We find that inelastic channels allow better sensitivity than elastic counterparts in all detectors for both heavy and light mediators across the whole range of DM masses considered - with the exception of DUNE.
For $m_{A^\prime}=100$ MeV and $m_\chi \lesssim 0.1$ MeV, DUNE's elastic sensitivity is stronger than inelastic.
This is due to a combination of factors.
With decreasing mass, the inelastic DM cross section drops faster than the elastic cross section and the available phase space also decreases.
This affects all detectors, however given carbon's much larger GT strength (relative to argon), the inelastic channel remains dominant in carbon based detectors but not argon ones.

JUNO achieves the best overall sensitivity (with the inelastic channel) over almost the entire DM mass range.
The strong GT transition and light mass of the carbon nucleus imply a higher signal rate per unit detector mass relative to both oxygen and argon targets.
DUNE has twice the exposure as JUNO but lower GT strengths than JUNO.
The number of atoms in DUNE is less than JUNO since argon atomic mass is higher than carbon.
Thus, DUNE has the second best sensitivity.
Borexino ranks the third due to the signal rate of carbon.
Super-K and Hyper-K are at disadvantage in sensitivity search because of the low $^{16}$O branching ratio.
Hyper-K is relatively better than due to its large detector mass.
Given that Borexino and Super-K have accrued significant exposures, our exploratory result motivates a revisit of the data which could ultimately prove to be the strongest constraint on this parameter space to-date.

The ``U'' shape of the inelastic sensitivity curves is caused by the opposite behavior of the differential of $\Phi_\chi$ and $\sigma^{\mathrm{inel}}_{\chi N}$ with respect to $m_\chi$, i.e., $\Phi_\chi$ decreases whereas $\sigma^{\mathrm{inel}}_{\chi N}$ increases in increasing $m_\chi$.
Therefore, there exists a sweet spot where the product $\sigma^{\mathrm{inel}}_{\chi N} \, \Phi_\chi$ is maximized, which corresponds to the lowest point of $\epsilon$ in Fig.~\ref{fig:sen}.

\begin{figure*}[t]
    \centering
    \includegraphics[width=2\columnwidth]{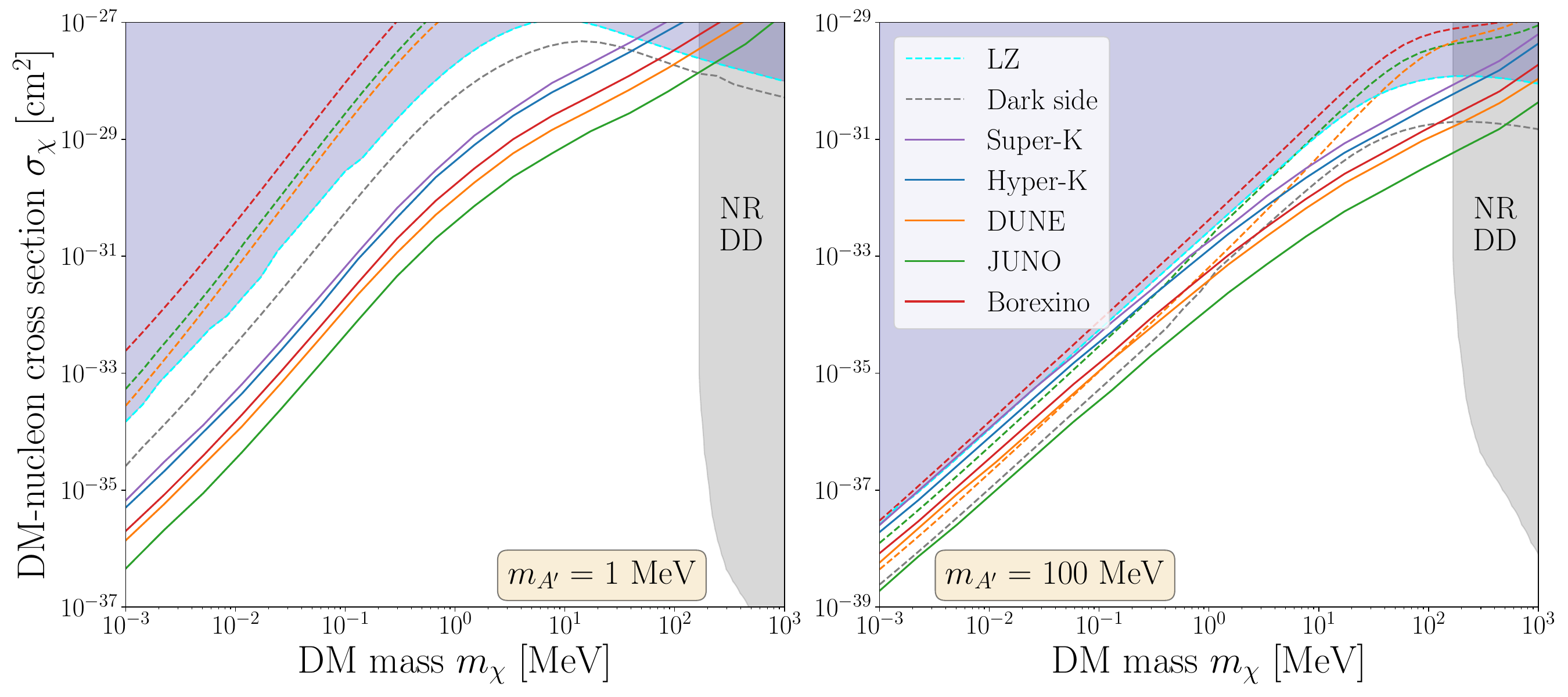}
    \caption{Similar to Fig.~\ref{fig:sen}, but for DM-nucleon scattering cross-section $\sigma_\chi$. The grey region is the constraints from non-relativistic direct detection, CRESST-III~\cite{CRESST:2019jnq}.
    }
    \label{fig:sen_xsec}
\end{figure*}

Finally, for ease of comparison, we interpret our results in terms of the elastic DM-nucleon cross-section $\sigma_\chi$ in the non-relativistic and heavy-mediator limit, showing them in Fig.~\ref{fig:sen_xsec}.
The CRDM flux with $\sigma_\chi \gtrsim 10^{-27}~{\rm cm}^2$ may experience significant attenuation, requiring more careful analyses. The most stringent constraints from standard direct detection experiments in this region come from CRESST-III~\cite{CRESST:2019jnq} (gray-shaded regions), they do not extend below $m_\chi=200$~MeV however.

\medskip

\noindent {\bf Conclusions.}
In conclusion, the inelastic nuclear scattering channel has several advantages.
First, it requires no low thresholds since the nuclear-line energy is $\sim \mathcal{O}$(10) MeV.
Second, it enjoys a statistical advantage due to lower backgrounds.
They allow us to leverage the sensitivity of large-volume neutrino detectors for probing light DM.
We have shown that the projected sensitivity reaches in the inelastic scattering channel are better than those in the elastic channel for DM masses in the range of keV $-$ GeV for all neutrino experiments considered.
These projections can far exceed the current best bounds from the elastic channel at LZ.

\begin{acknowledgments}
The work of BD, WH, and DK is supported in part by the U.S.~Department of Energy Grant DE-SC0010813.
The work of JLN and ISA is supported by the Australian Research Council through the ARC Centre of Excellence for Dark Matter Particle Physics, CE200100008.
The work of JCP is supported by the National Research Foundation of Korea grant funded by the Korea government(MSIT) (NRF-2019R1C1C1005073, RS-2024-00356960).
We thank Koun Choi for the valuable discussion.
\end{acknowledgments}

\bibliography{refs}

\end{document}


\begin{flushright}
MI-HET-827
\end{flushright}
\title{Supplement material: Prospects for Light Dark Matter Searches at Large-Volume Neutrino Detectors
}

\author{Bhaskar Dutta}
\email{dutta@physics.tamu.edu}
\affiliation{Mitchell Institute for Fundamental Physics and Astronomy$,$ Department of Physics and Astronomy$,$ \\Texas A\&M University$,$ College Station$,$ TX 77843$,$ USA}

\author{Wei-Chih Huang}
\email{s104021230@tamu.edu}
\affiliation{Mitchell Institute for Fundamental Physics and Astronomy$,$ Department of Physics and Astronomy$,$ \\Texas A\&M University$,$ College Station$,$ TX 77843$,$ USA}

\author{Doojin Kim}
\email{doojin.kim@usd.edu}
\affiliation{Mitchell Institute for Fundamental Physics and Astronomy$,$ Department of Physics and Astronomy$,$ \\Texas A\&M University$,$ College Station$,$ TX 77843$,$ USA}
\affiliation{Department of Physics$,$ \\University of South Dakota$,$ Vermillion$,$ SD 57069$,$ USA}

\author{Jayden~L.~Newstead}
\email{jnewstead@unimelb.edu.au}
\affiliation{ARC Centre of Excellence for Dark Matter Particle Physics$,$ School of Physics$,$ \\The University of Melbourne$,$ Victoria 3010$,$ Australia}

\author{Jong-Chul Park}
\email{jcpark@cnu.ac.kr}
\affiliation{Department of Physics and Institute of Quantum Systems (IQS)$,$ \\Chungnam National University$,$ Daejeon 34134$,$ Republic of Korea}

\author{Iman Shaukat Ali}
\email{ishaukatali@student.unimelb.edu.au}
\affiliation{ARC Centre of Excellence for Dark Matter Particle Physics$,$ School of Physics$,$ \\The University of Melbourne$,$ Victoria 3010$,$ Australia}

\maketitle

\begin{figure}[tbh]
    \centering
    \includegraphics[width=\columnwidth]{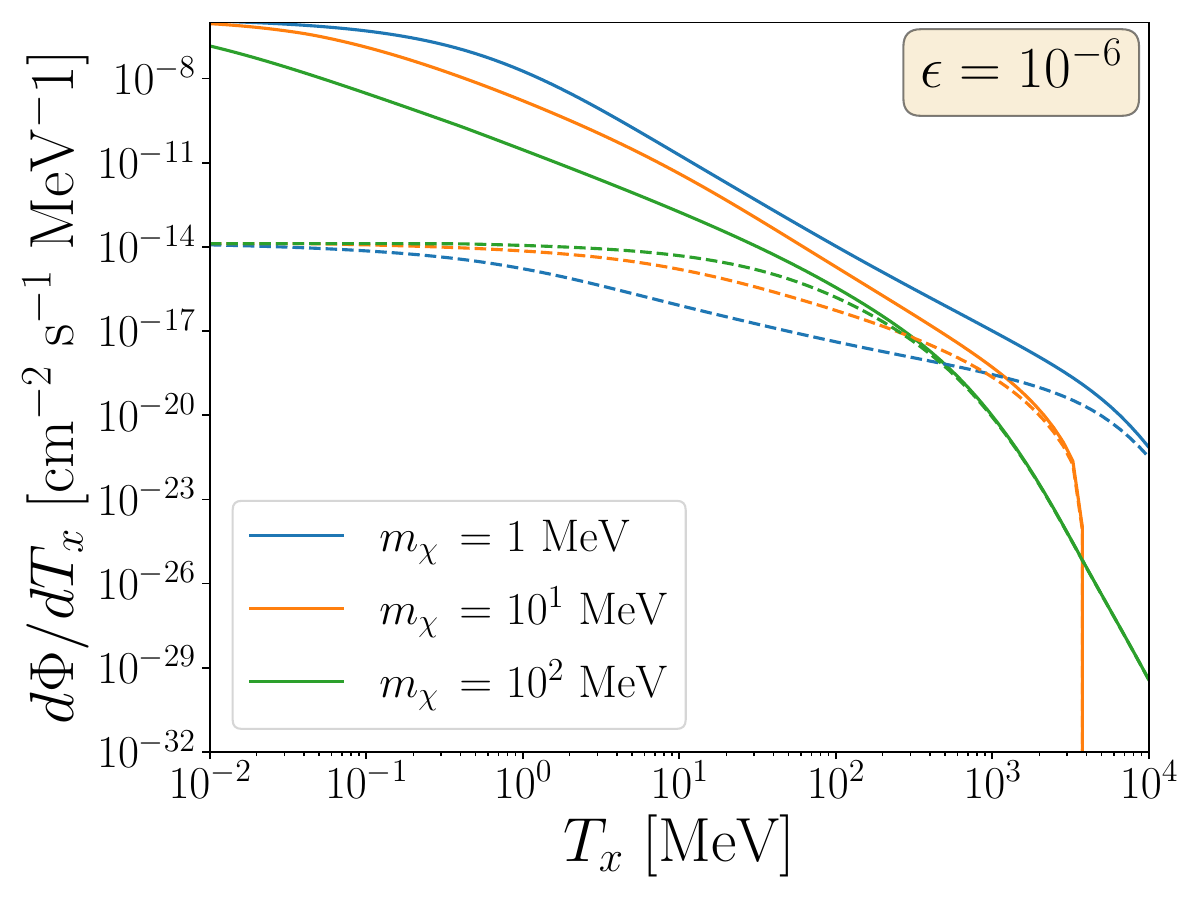}\\
    \caption{CRDM flux for different $m_\chi$ with fixed $\epsilon=10^{-6}$ and $m_{A^\prime}=1$ MeV (solid) and 100 MeV (dashed).}
    \label{fig:flux}
\end{figure}

\section{CRDM flux}
In Fig.~\ref{fig:flux}, we show the CRDM flux for different $m_\chi$ with fixed $\epsilon$ and $m_{A^\prime}$.
We consider the DM boosting via the elastic channel only since the cross-section for this process dominates over the inelastic processes we considered.

\section{GT strengths}
\begin{figure}[h]
    \centering
    \includegraphics[width=\columnwidth]{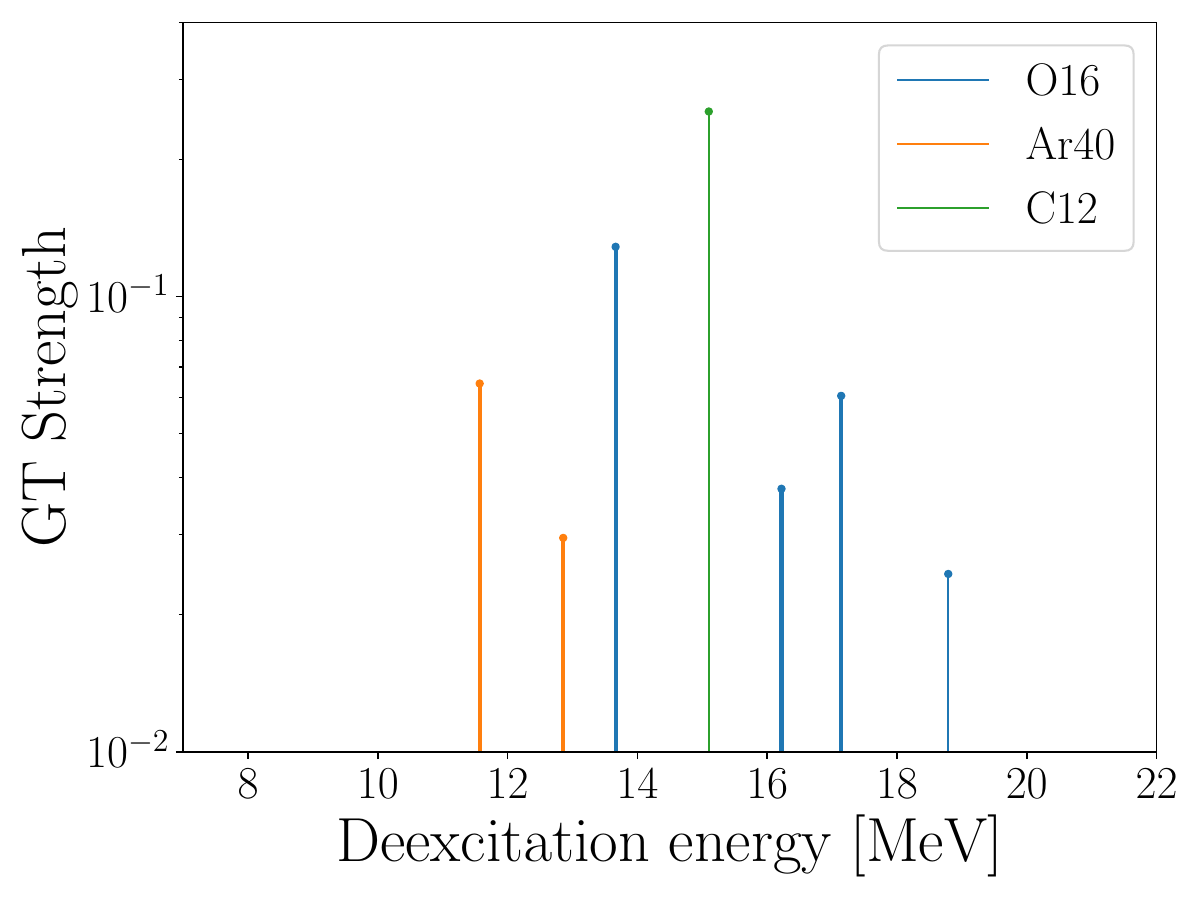}
    \caption{The GT strengths for $^{40}$Ar (calculated by BIGSTICK and used for DUNE), $^{16}$O (given by measurements and used for Super-K/Hyper-K), and $^{12}$C (given by measurements and used for JUNO/Borexino).
    }
    \label{fig:strengths}
\end{figure}

A summary of the GT transition strengths $|\langle J_f|| \sum_{i=1}^A \frac{1}{2}\hat{\sigma_i} \hat{\tau_0}|| J_i\rangle|^2$ for the targets used in this work is shown in Fig.~\ref{fig:strengths}.

\section{Nuclear shell model}
\label{sec:nuclear}
In large-scale shell model calculations, it is necessary to reduce
the computational burden by truncating the many-body Hilbert space of the nucleus.
This is done by considering a subset of states that have a frozen
core and a valence space.
Depending on the particular nuclei of interest, one restricts the valence space of the protons and neutrons based on the
outermost orbits they lie in.
For $^{40}$Ar, the valence protons lie within $sd$ orbits ($0d_{5/2}$, $1s_{1/2}$, $0d_{3/2}$) while the valence neutrons are in the $pf$ ($0f_{7/2}$, $1p_{3/2}$, $0f_{5/2}$, $1p_{1/2}$) orbits.
Attempting the calculation on the entire $sdpf$ valence space is far too computationally heavy.
By separately restricting protons in the $sd$ orbits and neutrons in the $pf$ orbits, the computational cost is made manageable.

\bibliography{refs}